\documentstyle[12pt,psfig]{article}
\newlength{\dinwidth} 
\newlength{\dinmargin}
\newcommand{\m}      {\:\rm m}
\newcommand{\mm}      {\:\rm mm}

\newcommand{\TeV}      {{\:\rm TeV}}
\newcommand{\GeV}      {\:\rm GeV}
\newcommand{\MeV}      {\:\rm MeV}

\newcommand{\counts}      {\:\rm counts}
\newcommand{\mrad}      {\:\rm mrad}
\newcommand {\pom}  {I\hspace{-0.2em}P}
\newcommand {\po}  {I\hspace{-0.3em}P}
\newcommand {\alphapom} {\mbox{$\alpha_{\pom}$}}
\newcommand {\alphapomp} {\mbox{$\alpha'_{\pom}$}}
\newcommand {\reg}  {I\hspace{-0.2em}R}
\newcommand {\re}  {I\hspace{-0.3em}R}
\newcommand {\alphareg} {\mbox{$\alpha_{\reg}$}}

\begin{document}
\renewcommand{\thefootnote}{\arabic{footnote}}
\begin{titlepage}

\vspace{2cm}
\begin{center}

{\LARGE
Study of Photon Dissociation in\\
\vspace{.6cm}
Diffractive Photoproduction at HERA\\}

\vspace{1cm}
{\large ZEUS Collaboration}

\vspace{1cm}
\end{center}

\vspace{3.5cm}

\centerline{\bf Abstract}
\vspace{.2cm}

Diffractive dissociation of quasi--real photons
at a photon--proton centre of mass energy of \mbox{$W \approx 200\GeV$}
is studied with the ZEUS detector at HERA.
The process under consideration is \mbox{$\gamma p\rightarrow XN$},
where \mbox{$X$} is the diffractively dissociated photon system of mass \mbox{$M_X$} and
\mbox{$N$} is either a proton or a nucleonic system with mass \mbox{$M_N<2\GeV$}.
The cross section for this process 
in the interval \mbox{$3<M_X<24\GeV$} relative to the total
photoproduction cross section was measured to be
\mbox{$\sigma^{partial}_{D}/\sigma_{tot} = 6.2 \pm 0.2 (stat) \pm 1.4 (syst)\%$}.
After extrapolating this result 
to the mass interval of \mbox{$\m_{\phi}^2<M_X^2<0.05W^2$}
and correcting it for proton dissociation,
the fraction of the total cross section attributed to
single diffractive photon dissociation, \mbox{$\gamma p\to Xp$}, is found to be
\mbox{$\sigma_{SD}/\sigma_{tot} = 13.3 \pm 0.5 (stat)\pm 3.6(syst)\%$}.
The mass spectrum of the dissociated photon system
in the interval \mbox{$8<M_X<24\GeV$} can be described
by the triple pomeron (\mbox{$\po\po\po$}) diagram with an effective 
pomeron intercept of \mbox{$\alphapom(0)=1.12\pm 0.04(stat) \pm 0.08(syst)$}.
The cross section for photon dissociation in the range \mbox{$3<M_X<8\GeV$} is
significantly higher than that expected from the triple pomeron amplitude 
describing the region \mbox{$8<M_X<24\GeV$}.
Assuming that this discrepancy is due to a 
pomeron--pomeron--reggeon (\mbox{$\po\po\re$}) term, its contribution 
to the diffractive cross section in the interval \mbox{$3<M_X<24\GeV$} is estimated
to be \mbox{$f_{\pom\pom\reg}=26\pm 3(stat) \pm 12(syst)\%$}.

\end{titlepage}

\newpage
{
\parindent0.cm                                                                                     
\parskip0.3cm plus0.05cm minus0.05cm                                           
                   
\def\3{\ss}                                                                     
\pagenumbering{Roman}                                                           
\footnotesize

%
%
%
%
%
                                                   %
\begin{center}                                                                                     
{                      \Large  The ZEUS Collaboration              }                               
\end{center}                                                                                       
  J.~Breitweg,                                                                                     
  M.~Derrick,                                                                                      
  D.~Krakauer,                                                                                     
  S.~Magill,                                                                                       
  D.~Mikunas,                                                                                      
  B.~Musgrave,                                                                                     
  J.~Repond,                                                                                       
  R.~Stanek,                                                                                       
  R.L.~Talaga,                                                                                     
  R.~Yoshida,                                                                                      
  H.~Zhang  \\                                                                                     
 {\it Argonne National Laboratory, Argonne, IL, USA}~$^{p}$
% ...
\newpage                                                                                           
% ...
$^{\    2}$ now at Univ. of Salerno and INFN Napoli, Italy \\                                      
$^{\    3}$ now at Univ. of Crete, Greece \\                                                       
% ...
$^{\    8}$ now at Dongshin University, Naju, Korea \\                                             
% ...
$^{  23}$ now at ATLAS Collaboration, Univ. of Munich \\                                           
$^{  24}$ now at Clinical Operational Research Unit,                                               
University College, London\\                                                                       
% ...
                                                           %
                                                           %
\newpage   
                                                           %
                                                           %
\begin{tabular}[h]{rp{14cm}}                                                                       
$^{a}$ &  supported by the Natural Sciences and Engineering Research                               
          Council of Canada (NSERC)  \\                                                            
$^{b}$ &  supported by the FCAR of Qu\'ebec, Canada  \\                                            
$^{c}$ &  supported by the German Federal Ministry for Education and                               
          Science, Research and Technology (BMBF), under contract                                  
          numbers 057BN19P, 057FR19P, 057HH19P, 057HH29P, 057SI75I \\                              
$^{d}$ &  supported by the MINERVA Gesellschaft f\"ur Forschung GmbH,                              
          the German Israeli Foundation, and the U.S.-Israel Binational                            
          Science Foundation \\                                                                    
$^{e}$ &  supported by the German Israeli Foundation, and                                          
          by the Israel Science Foundation                                                         
  \\                                                                                               
$^{f}$ &  supported by the Italian National Institute for Nuclear Physics                          
          (INFN) \\                                                                                
$^{g}$ &  supported by the Japanese Ministry of Education, Science and                             
          Culture (the Monbusho) and its grants for Scientific Research \\                         
$^{h}$ &  supported by the Korean Ministry of Education and Korea Science                          
          and Engineering Foundation  \\                                                           
$^{i}$ &  supported by the Netherlands Foundation for Research on                                  
          Matter (FOM) \\                                                                          
$^{j}$ &  supported by the Polish State Committee for Scientific                                   
          Research, grant No.~115/E-343/SPUB/P03/120/96  \\                                        
$^{k}$ &  supported by the Polish State Committee for Scientific                                   
          Research (grant No. 2 P03B 083 08) and Foundation for                                    
          Polish-German Collaboration  \\                                                          
$^{l}$ &  partially supported by the German Federal Ministry for                                   
          Education and Science, Research and Technology (BMBF)  \\                                
$^{m}$ &  supported by the German Federal Ministry for Education and                               
          Science, Research and Technology (BMBF), and the Fund of                                 
          Fundamental Research of Russian Ministry of Science and                                  
          Education and by INTAS-Grant No. 93-63 \\                                                
$^{n}$ &  supported by the Spanish Ministry of Education                                           
          and Science through funds provided by CICYT \\                                           
$^{o}$ &  supported by the Particle Physics and                                                    
          Astronomy Research Council \\                                                            
$^{p}$ &  supported by the US Department of Energy \\                                              
$^{q}$ &  supported by the US National Science Foundation \\                                       
\end{tabular}                                                                          
}

\newpage

\pagenumbering{arabic}

\setcounter{page}{1}

\normalsize

\section{Introduction}
Interactions of real photons with protons 
at high energy bear many similarities to hadronic interactions.
This can be understood in the framework of
the vector meson dominance model (VDM) \cite{Sakurai,Bauer}, in which
the photon is assumed to fluctuate into a virtual 
vector meson (\mbox{$\rho , \omega , \phi$}) prior to the
interaction with the proton.
The resulting collisions are expected to exhibit all the 
characteristics of a hadron--hadron reaction, including 
diffraction.  Diffractive processes at high energies
are generally characterized by an exponential suppression of the
squared four momentum (\mbox{$t$}) transferred between the colliding
particles and a weak energy dependence of the cross section.
The colliding hadrons may emerge intact from the
interaction (elastic scattering).  Alternatively, one or both may
be excited into more massive states -- single or double diffractive
dissociation, respectively.  In all cases the hadronic final
state is characterized by the presence of two groups of particles
separated in rapidity.

The elastic and total cross sections for
hadron--hadron collisions at high centre of mass (c.m.) energies
have been successfully described in the Regge picture in terms
of the exchange of two dominant trajectories: 
the pomeron and the reggeon~\cite{DL-sigtot}.
The data from fixed target photoproduction experiments~\cite{Bauer}
combined with the recent measurements
from HERA~\cite{H1-sigtot,ZEUS-sigtot,ZEUS-rho,H1-rho} 
also confirm the validity of this model
for the description of the total photoproduction cross section as well as for
the cross sections for the light vector meson production.

Regge theory in conjunction with Mueller's theorem~\cite{Mueller} allows the
modelling of single dissociation processes~\cite{Goulianos}.  
The measurements of \mbox{$pp$} and \mbox{$p\bar{p}$}
reactions up to the very high c.m. energies of \mbox{$W=1800 \GeV$}~\cite{CDF-mx} show
that the diffractive cross sections are dominated by
the triple pomeron amplitude.  
The values of the pomeron intercept extracted from
the shape of the dissociated mass spectra are consistent with those
obtained from the total and elastic cross sections~\cite{Field-Fox}.
The diffractive dissociation of real photons
has been previously studied at fixed target experiments 
reaching a c.m. energy of \mbox{$W \approx 14\GeV$}~\cite{Chapin}, where 
it was observed that the general properties of diffractive photoproduction
are similar to those of hadronic reactions.
It is thus of interest to test whether this similarity holds 
at higher energies.

In this paper we study the dependence of the cross section 
on the mass, \mbox{$M_X$},  of the dissociated photon system \mbox{$X$} for the
diffractive process \mbox{$\gamma p\to Xp$} .  The measurement was performed
with the ZEUS detector at the HERA collider using \mbox{$ep$} 
collisions in which the virtuality \mbox{$Q^2$} of the
exchanged photon is smaller than \mbox{$0.02\GeV^2$} and 
\mbox{$W \approx 200\GeV$}.  A similar analysis has been performed recently by
the H1 collaboration~\cite{H1-mx}.

The paper is organized as follows. After a brief review of the basic 
concepts of
the triple Regge phenomenology (section \ref{s:regge}), 
we describe the experimental
setup, the trigger and the event selection criteria (sections \ref{s:ex_setup},
\ref{s:trigger} and \ref{s:ev_sel}). 
The Monte Carlo (MC)
models used for the acceptance corrections are described in section \ref{s:mc},
while section \ref{s:mx_rec} 
contains the presentation of the method used to reconstruct \mbox{$M_X$}.
In section \ref{s:rap_gap} we describe the measurement of 
the \mbox{$M_X$} spectrum in events
corresponding to the processes \mbox{$\gamma p\to XN$}, 
where \mbox{$N$} is either a
proton or a nucleonic system with mass \mbox{$M_N<2\GeV$}.
In order to suppress the contribution from nondiffractive
photoproduction processes only the events with a gap in
the rapidity distribution of final state hadrons (rapidity--gap events)
were included.  The subtraction of the remaining contamination from
nondiffractive processes and the
correction for detector effects were performed
using a MC simulation technique.  The analysis of the \mbox{$M_X$}
spectrum in the framework of Regge theory is described in section 
\ref{s:anal_rap_gap}.
To test the sensitivity of the results to the model assumptions made in
this study an alternative
analysis of the same data sample was performed.  No
rapidity--gap was required and the distinction between the different
processes was performed solely on the basis of the shape of the \mbox{$M_X$}
spectrum.  This alternative analysis is described in section 
\ref{s:lnMx2_anal}, which also contains
a comparison between the results of the two methods.
The paper concludes, in section \ref{s:concl},
with a comparison of the results with
other experiments and with the expectations from Regge phenomenology. 

\section{Triple Regge model}
\label{s:regge}
\begin{figure}[tb]
\centerline{\hbox{
\psfig{figure=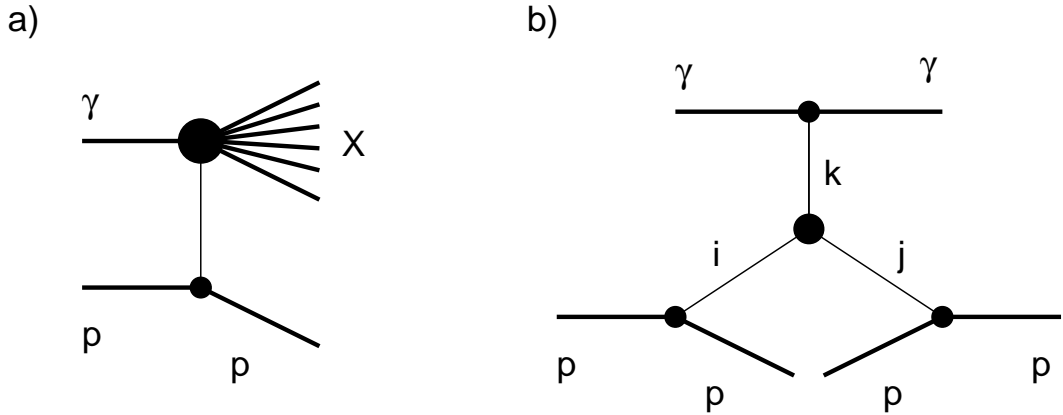,width=14cm}}}
\bf\caption{\it 
The photoproduction processes related by Mueller's theorem:
\mbox{$a)$} the inclusive reaction \mbox{$\gamma p\rightarrow Xp$} 
and \mbox{$b)$} the triple Regge diagram of the three body scattering 
\mbox{$\gamma pp\rightarrow \gamma pp$}.
}
\label{f:triple-Regge}
\end{figure}

The Regge model describes particle interactions in terms of the exchange of 
trajectories.  Three trajectories are of primary importance, namely
the pomeron, \mbox{$\alphapom(t)=1.08 + 0.25\cdot t$},
the reggeon, \mbox{$\alphareg(t) = 0.45 + t$}
and the pion \mbox{$\alpha_{\pi}(t) = 0 + t$}.  The
trajectory parameters are not given by the model but are determined from
data~\cite{Field-Fox,DL-sigtot}.
Mueller's theorem~\cite{Mueller} relates 
the inclusive cross section for the photoproduction reaction
\mbox{$\gamma p\rightarrow Xp$}  (figure \ref{f:triple-Regge}a) 
to the forward amplitude of the three body hadronic process
\mbox{$\gamma pp\rightarrow \gamma pp$}.
If \mbox{$M_X^2$}, \mbox{$W^2/M_X^2$} and \mbox{$W^2/|t|$} 
are large~\cite{Collins}, the triple Regge
diagram shown in figure \ref{f:triple-Regge}b is expected to dominate
the three body amplitude.  The triple Regge diagram predicts
the following behaviour of the cross section~\cite{Collins}:
\begin{equation}
\frac{d^2 \sigma}{d t d M_{X}^2} =
 \left(\frac{1}{W^2}\right)^2\cdot 
 \sum_{ijk} G_{ijk}(t) 
 \left(\frac{W^2}{M_X^2}\right)^{\alpha_i(t)+\alpha_j(t)}
  {M_X^2}^{\alpha_k(0)},
\label{e:tripple-Regge}
\end{equation}
where the indices \mbox{$i,j,k$} denote the Regge trajectories.
The effective coupling strength, \mbox{$G_{ijk}(t)$},
is not predicted by the model and must be determined from the experimental 
data. In the process depicted in figure \ref{f:triple-Regge}b only the
trajectories denoted by \mbox{$i$} and \mbox{$j$} are exchanged between the
colliding particles and they carry the four momentum squared \mbox{$t$}.
The $k$ trajectory is related, via the optical theorem, to the
probability that either $i$ or $j$ couple to the photon~\cite{Collins-Martin}.

Diffractive processes are attributed to the exchange of
the pomeron.  In the triple Regge regime two diagrams are of
primary importance: \mbox{$ijk=\po\po\po$} and
\mbox{$\po\po\re$}~\cite{Field-Fox}.  The former triple pomeron term 
leads to an inclusive cross section falling with \mbox{$M_X^2$} 
approximately as
\mbox{$d \sigma/d M_{X}^2 \propto 1/M_X^2$}.  
The \mbox{$\po\po\re$} contribution
is important only at lower diffractive masses
as it gives a steeper \mbox{$M_X^2$} dependence of the form
\mbox{$d \sigma/d M_{X}^2 \propto (1/{M^2_X})^{1.5}$}.

A number of other triple Regge terms have been found to give a
contribution to the inclusive cross sections in hadron--hadron
reactions~\cite{Field-Fox}, namely:  \mbox{$ijk=\re\re\po$} describing the
reggeon exchange and \mbox{$ijk=\pi\pi\po$} and \mbox{$\pi\pi\re$} 
describing the exchange of a pion trajectory.  
However these terms give contributions which are negligible
at low \mbox{$M_X$} and become comparable to diffractive pomeron
exchange only at \mbox{$M_X^2\approx 0.05 W^2$}~\cite{CDF-mx,Field-Fox}. 

In this paper only the processes due to pomeron exchange are
referred to as diffractive and are treated as signal.
The processes due to reggeon and pion exchange are called
nondiffractive and are considered backgrounds.

\section{Experimental setup}
\label{s:ex_setup}
The analysis is based on data collected in 1994 with the ZEUS detector.
HERA operated at a positron energy of \mbox{$27.5\GeV$}
and a proton energy of \mbox{$820\GeV$}, with 153 colliding
bunches.  In addition 15 positron and 17 proton bunches were left unpaired
for background studies. 

A detailed description of the ZEUS detector may be found elsewhere   
\cite{status93,ZEUS-description}.
Here, a brief description of the detector components most relevant
for this analysis is given. 
Throughout this paper the standard ZEUS coordinate system is used, 
which has its origin at the nominal interaction point.
The $Z$--axis points in the direction of the proton beam,
called the forward direction, and the $X$--axis 
points towards the centre of the HERA ring.

For the energy measurement the high resolution 
depleted--uranium scintillator calorimeter (CAL) is used~\cite{CAL}. 
It is divided
into three parts, forward (FCAL) covering the 
pseudorapidity~\footnote{Pseudorapidity \mbox{$\eta$} is evaluated from the relation
\mbox{$ \eta = - ln( tan( \theta / 2))$}, where
\mbox{$\theta$} is a polar angle calculated with respect to the proton beam
direction.} 
region \mbox{$4.3>\eta>1.1$}, barrel (BCAL) covering the central region
\mbox{$1.1>\eta>-0.75$} and rear (RCAL) covering the backward region
\mbox{$-0.75>\eta>-3.8$}.  Holes of \mbox{$20 \times 20 {\rm\: cm}^{2}$} in
the centre of FCAL and RCAL accommodate the HERA beam
pipe.
Each of the calorimeter parts is subdivided into towers which in 
turn are segmented longitudinally into electromagnetic (EMC) and
hadronic (HAC) sections.  These sections are further subdivided
into cells, which are read out by two photomultiplier tubes.
Under test beam conditions, the energy resolution of the calorimeter 
was measured to be
\mbox{$\sigma_{E}/E = 0.18/\sqrt{E (\GeV)}$} for electrons and 
\mbox{$\sigma_{E}/E = 0.35/\sqrt{E (\GeV)}$} for hadrons.
The calorimeter noise, dominated by the uranium radioactivity, is in
the range \mbox{$15-19\MeV$} for EMC cells and \mbox{$24-30\MeV$} for HAC
cells.

The proton remnant tagger (PRT) is used to tag events with a rapidity--gap.
It consists of two layers of scintillation counters 
installed perpendicular to the beam at \mbox{$Z=5.15\m$}.
The two layers are separated by a \mbox{$2 \mm$} thick lead absorber.
Each layer is split into two halves along the $Y$--axis and
each half is independently read out by a photomultiplier tube.
The counters have an active area of
dimensions \mbox{$30\times 26\,{\rm cm^2}$} 
with a hole of \mbox{$6.0\times 4.5\,{\rm cm^2}$} 
at the centre to accommodate the HERA beam pipe.  The pseudorapidity
range covered by the PRT is \mbox{$4.3<\eta<5.8$}.

The luminosity monitor~\cite{LUMI} (LUMI) measures the rate of the
Bethe--Heitler process $e p \rightarrow e \gamma p$.
The detector consists of two lead--scintillator sandwich 
calorimeters, installed in the HERA tunnel.  The one at $Z=-35\m$ is
designed to detect positrons scattered at very small
angles and the one at $Z=-107\m$ measures the
photons emitted along the positron beam direction.
In this analysis, signals in the LUMI positron calorimeter were 
used to tag photoproduction
events with positrons scattered at angles up to about $5\mrad$ with
respect to the positrons beam direction.   The LUMI positron calorimeter
was also used to measure the energy of the scattered positron, $E'_{e}$, 
and derive the energy of the exchanged quasi--real photon, $E_{\gamma}$,
through the relation \mbox{$E_{\gamma}=E_{e}-E'_{e}=27.5{\GeV} -E'_{e}$}.

The leading proton spectrometer (LPS)~\cite{LPS} 
detects charged particles scattered at 
small angles and carrying a substantial fraction of the incoming
proton momentum.  These particles remain in the beam pipe and their trajectory
is measured by a system of position sensitive silicon micro--strip detectors
installed very close to the proton beam at $Z=63.0\m$, $81.2\m$ and $90.0\m$.
The track deflection induced by the magnets in the proton beam line
is used for the momentum analysis of the scattered proton.

\section{Trigger}
\label{s:trigger}
ZEUS uses a three level trigger system.  At the first level
 a  coincidence between signals in the LUMI positron
calorimeter and in the rear part of the uranium calorimeter was required.
The small angular acceptance of the LUMI positron calorimeter
implied that the virtuality of the exchanged photon was 
\mbox{$Q^{2}< 0.02 \GeV^{2}$}.
The uranium calorimeter trigger required a measured energy deposit in the
RCAL EMC section of more than \mbox{$464\:{\rm MeV}$} 
(excluding the towers immediately adjacent to the beam pipe) 
or \mbox{$1250\:{\rm MeV}$} (including those towers).

The second and the third trigger levels were mainly used to reject
beam related background.  Parts of the data stream were
prescaled in order to reduce the high event rate resulting from
the large photoproduction cross section.  

\section{Selection of photoproduction events}
\label{s:ev_sel}
The sample of events satisfying the photoproduction
trigger and used in this study 
consisted of 103k events from a luminosity of \mbox{$0.7{\:\rm pb^{-1}}$}.  
In the offline analysis the energy 
of the scattered positrons measured in the LUMI calorimeter was restricted to
the range \mbox{$12<E'_{e}<18\GeV$}, thereby limiting the \mbox{$\gamma p$} c.m. energy
to the interval of \mbox{$176<W<225\GeV$}.

\subsection{Calorimeter noise suppression and trigger correction}
The offline data sample contained a small number of events
accidentally accepted by the online trigger 
because of a photomultiplier discharge or calorimeter noise contributing to an
energy sum sufficient to exceed the trigger threshold.  
Thus, in the offline analysis, each 
event was subject to a two step trigger correction procedure.
In the first step a noise suppression algorithm was
applied to the CAL data.
All the EMC (HAC) cells with energy below \mbox{$60\MeV$} \mbox{$(110\MeV)$}
were excluded from the data.
For isolated cells the thresholds were increased to \mbox{$80\MeV$} \mbox{$(140\MeV)$}. 
Isolated cells were also excluded if 
they corresponded to one of the known noisy readout channels or
if the imbalance between the
two corresponding PM tubes was too large, indicating a noise pulse.
This noise suppression algorithm was developed using
events collected with a random trigger.   In the second step
the corrected CAL energies were used to reevaluate the trigger
decision.
The photoproduction events that failed the offline reconstructed
trigger were not used in the analysis.

\subsection{Statistical background subtraction}
The remaining contamination of the offline sample was of two types:
the \mbox{$e$}--gas and the coincidence background. 
The contamination of the data sample from \mbox{$e$}--gas background 
was on average below \mbox{$0.5\%$}, and
concentrated in the sample of events characterized by low energy deposits
in CAL.  It was statistically subtracted using events 
from beam crossings where the positron bunch was unpaired.
Another type of background is due to  
events with accidental coincidence of
the bremsstrahlung process \mbox{$(ep \rightarrow e \gamma p)$}
triggering the LUMI positron calorimeter and
some activity in the main detector satisfying the RCAL trigger.
The contamination from this type of background is \mbox{$2\%$} on average.
It was subtracted statistically exploiting the fact that
a large fraction of these background events could be identified
since the energy deposits in the LUMI positron and
photon calorimeters summed up to the positron beam energy. 
The identified background events were
included with negative weights into all of the distributions
in order to compensate for the background events where the 
bremsstrahlung photon was not detected.
A detailed description of the statistical background 
subtraction method may be found in \cite{ZEUS-sigtot,Burow}.

\section{Monte Carlo simulation}
\label{s:mc}
\subsection{Models}
The diffractive photon dissociation process, \mbox{$\gamma p \rightarrow X p$},
was simulated with a MC program based on the 
Nikolaev--Zakharov~\cite{NZ} (NZ) model 
interfaced to the Lund fragmentation scheme \cite{Solano}.  
For the study of systematic uncertainties the same process 
was simulated with the EPSOFT~\cite{EPSOFT} program developed in the 
framework of 
HERWIG 5.7 \cite{HERWIG}.  EPSOFT models the diffractive dissociation
as a soft hadronic collision of the photon with the pomeron.
The particle multiplicities and the momenta of the hadrons
transverse to the photon--pomeron collision axis are 
simulated using  parameterizations of existing diffractive 
data~\cite{UA4,ZEUS-pt},
while the longitudinal momenta are generated uniformly in phase space.
Initially, the cross sections assigned to the events generated by 
both of these MC programs were
consistent with the triple pomeron relation assuming a 
pomeron intercept \mbox{$\alphapom(0)=1.08$}.  
For the final analysis they were iteratively reweighted 
so as to give the best description of the measured distributions,
notably the reconstructed mass spectrum (see sec.~\ref{s:mx_rec}).

Elastic production of vector mesons,
\mbox{$\gamma p \rightarrow V p$} with 
\mbox{$V\equiv \rho^\circ,\, \omega,\,  \phi$},
and the diffractive processes involving the dissociation of the proton,
\mbox{$\gamma p \rightarrow V N, X N$}, were simulated using EPSOFT.
In the latter case the cross section calculations relied on 
parameterizations of the \mbox{$pp \rightarrow pp,pN$} data~\cite{Goulianos}.

Soft, nondiffractive collisions of the proton with hadronic fluctuations of 
the photon were also generated using the EPSOFT program.  
The particle multiplicities and 
the transverse momenta of the hadrons were
simulated using parameterizations of the hadron--hadron 
data~\cite{had-mult-pt}
tuned to describe also the ZEUS data~\cite{ZEUS-pt}.  
The longitudinal momenta
were generated uniformly in phase space.
The effect of leading baryon production was simulated in EPSOFT
in accordance with results from \mbox{$p\bar{p}$} data.  
In the limit where the momentum of
the leading baryon is close to that of the initial proton,
i.e. where the triple Regge approach applies,
the EPSOFT simulation gives results consistent with the combination of
the reggeon and pion exchanges.
The soft nondiffractive \mbox{$\gamma p$} interactions from EPSOFT were
enriched with hard, direct and resolved subprocesses 
simulated using HERWIG 5.7.
The lower cut--off on the transverse momentum of the
final--state partons, \mbox{$p_{T min}$}, was chosen to be \mbox{$3 \GeV$}.
For the parton densities of the colliding particles, the GRV--LO \cite{GRV}
photon and MRSD\mbox{$'$}\_ \cite{MRS} proton parameterizations were used.  
To cross check the sensitivity of the results to the nondiffractive model
a sample of events generated with the multipartonic
interaction option of PYTHIA 5.7~\cite{PYTHIA} was used.

All of the generated MC events were processed through the 
ZEUS detector simulation program based on GEANT and run 
through the same ZEUS reconstruction chain as the data.
The events were then subject to the same CAL noise suppression algorithm 
and trigger requirements as the data.

\subsection{Combination of the different MC samples}
\label{ss:mc_comb}
The different MC samples corresponding to the subprocesses discussed above 
were combined.
In the first step the MC samples corresponding to the
soft and hard nondiffractive components were
combined with relative normalizations giving the best description of
the measured transverse momentum distribution of charged tracks.
In the next step 
the relative contribution of the diffractive and nondiffractive components
was adjusted so as to reproduce the ratio between the number of 
events with no hits in the PRT and \mbox{$8<M_{X\:rec}<20\GeV$} 
(see sections \protect{\ref{ss:prt-rap-gap}} and \protect{\ref{s:mx_rec}}) 
and the number of events with total CAL energy \mbox{$E_{tot}>60\GeV$} in the data.
The former data sample is dominated by diffractive processes 
with dissociated photon mass far from the region of low \mbox{$M_{X}$} resonances,
while the latter consists mainly of nondiffractive events.
In the MC simulation, the ratio of these two channels 
depends slightly on the characteristics of the simulated events.
Therefore, this normalization procedure was performed independently for
all the combinations of the MC models used.
In all cases the results were consistent with the
corresponding ratios between the measured photoproduction cross 
sections~\cite{H1-sigtot,ZEUS-sigtot}.
The contribution from the vector meson
production process was set to \mbox{$15\%$} of all the photon--proton
interactions, as inferred from the HERA 
measurements~\cite{H1-sigtot,ZEUS-sigtot,ZEUS-rho,H1-rho}.

\begin{figure}[tb]
  \centerline{\hbox{ \psfig{figure=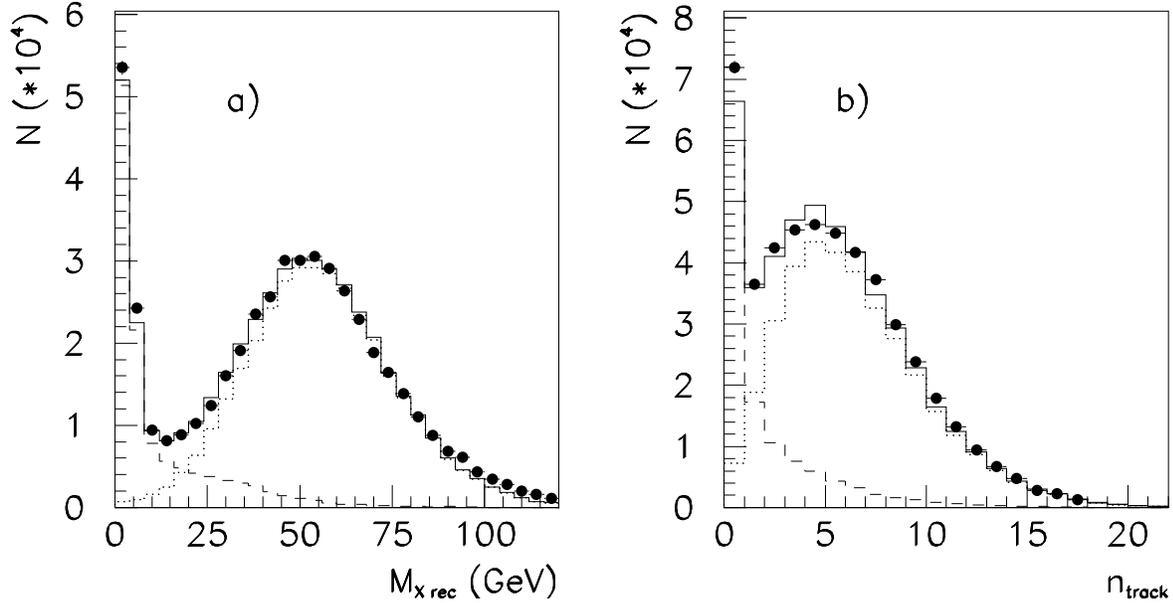}}}
\bf\caption {\it 
The distribution of a) the invariant mass of hadrons measured in 
the calorimeter (see sec.~\protect{\ref{s:mx_rec}})
and b) the multiplicity of charged tracks detected in the
region of \mbox{$-1.5<\eta<1.5$}.  
The data are shown as points and the result
of the MC simulation as a solid line.
The diffractive and the nondiffractive components are
also shown separately as dashed and dotted histograms, respectively.  }
\label{f:epsoft_data}
\end{figure}

The MC models used in this analysis were subject to a careful
selection and tuning.  The parameters defining the shapes of the
hadronic final states were adjusted by comparing the
distributions of multiplicity, polar angles and transverse momenta
of charged tracks in the MC simulation to those measured~\cite{ZEUS-pt}.  
As a result the Monte Carlo model correctly describes the general
characteristics of  photoproduction events at 
\mbox{$W\approx 200\GeV$} and
there is good agreement between the data and the simulation
for all relevant kinematical variables.
An example is presented in figure~\ref{f:epsoft_data} 
which shows a comparison between data
and MC for the invariant mass of the hadronic system measured in the CAL
(see below) and the multiplicity of charged tracks measured in the
interval \mbox{$-1.5<\eta<1.5$}.

\section{Mass reconstruction}
\label{s:mx_rec}
In the kinematic region of diffractive
photoproduction at HERA the dissociated photon system is produced nearly at
rest in the laboratory system.
Therefore, most of the particles from the photon dissociation are
produced within the geometric acceptance of the CAL, and
\mbox{$M_X$} may be approximated by
the measured invariant mass of the hadronic system:
\begin{equation}
M_{X\:rec}  = \sqrt{E^{2}-P^{2}} \approx
\sqrt{(E-P_{Z}) \cdot (E+P_{Z})} =
\sqrt{2 E_{\gamma} \cdot (E+P_{Z}), }
\end{equation}
where \mbox{$E_{\gamma}$} is the energy of the exchanged photon and
\mbox{$E$} is the energy of the hadronic system observed in the CAL.
The total momentum of the hadronic system, \mbox{$P$}, 
approximately equals the longitudinal component, \mbox{$P_Z$}, as the
transverse component is very small in the case of photoproduction events.
The following formula was used for the mass reconstruction:
\begin{equation}
M_{X\:rec} \equiv a_1 \cdot
\sqrt{ 2 (E_{e}-E_{e'}) \cdot 
(\sum_{cond}E_i + \sum_{cond}E_i cos\theta_i )}+a_2.
\label{e:mx_rec}
\end{equation}
The energy of the scattered positron, \mbox{$E_{e'}$}, was measured in the
LUMI positron calorimeter.  
The quantities \mbox{$E_i$} and \mbox{$\theta_i$} denote the energy and the polar angle of
CAL condensates,
defined as groups of adjacent cells with total energy
of at least \mbox{$100\MeV$}, if all the cells belong to the EMC, or
\mbox{$200\MeV$} otherwise. 

In order to test the sensitivity of the \mbox{$M_X$} measurement to
low energy particles which suffer from larger energy losses in the
inactive material before entering the CAL, the whole
analysis was repeated using only condensates of at least \mbox{$200\MeV$}.
The difference in the result is used for the estimate of the systematic error.
The coefficients \mbox{$a_1$} and \mbox{$a_2$} correct for the effects of energy loss in
the inactive material 
and energy deposits below the threshold.
Their values, \mbox{$a_1=1.14$} and \mbox{$a_2=1.2\GeV$},
were selected so as to give the best estimate of
the true invariant mass in diffractive photon dissociation events
from the MC simulation.

\begin{figure}[tb]
\centerline{\hbox{
\psfig{figure=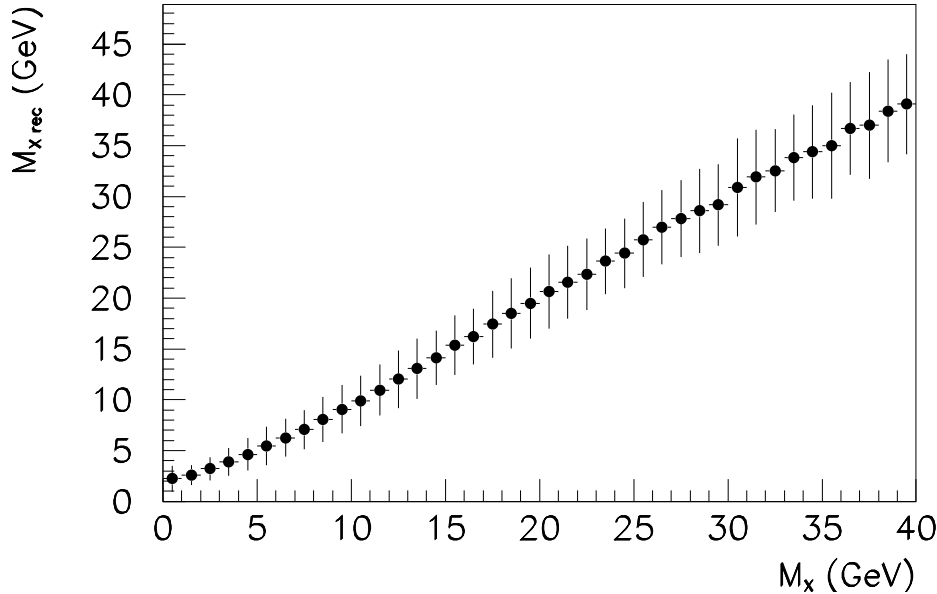}}}
\bf\caption{\it 
The relation between the 
generated and the reconstructed mass of the dissociated photon
in diffractive photoproduction events simulated using the EPSOFT MC program.
The error bars show the r.m.s. of the reconstructed mass.
}
\label{f:mx_scatter}
\end{figure}

Figure~\ref{f:mx_scatter}
illustrates the quality of the diffractive 
mass reconstruction in the events from the MC simulation.  
The masses in the range \mbox{$4<M_{X}<40\GeV$} are reconstructed with an
approximate resolution of \mbox{$\sigma(M_X)/M_X \approx 80\%/\sqrt{M_X}$} and
an offset  smaller than \mbox{$0.5\GeV$}.
The quality of the mass reconstruction has also been verified in the data
using the events where the scattered proton was measured in the LPS.  In
these events the invariant mass of the hadronic system was estimated from
the relation \mbox{$M_X^2\approx W^2\cdot (1-x_L)$}, where \mbox{$x_L=p'_p/p_p$} is the
fraction of the initial momentum retained by the scattered proton.  
The distribution of the difference between the mass reconstructed from CAL and
that estimated from the LPS
shows a gaussian peak corresponding to contained
events, i.e. where the entire $X$ system was detected in CAL, and
long tails due to events where some of the hadrons escaped detection
through the beam pipe hole.  For events with
\mbox{$4<M_{X}<45\GeV$} the centre of the peak was 
at \mbox{$0\pm 0.5\GeV$} confirming
that the \mbox{$M_X$} reconstruction using the calorimeter showed
no significant shifts.

At very low masses, \mbox{$M_{X}<2\GeV$}, the mass reconstruction in CAL
suffers from migrations towards higher
values of \mbox{$M_{X\:rec}$} due to the limited angular resolution of the
calorimetric measurement.  To reduce these migrations
an additional cut was applied which accepted only events with at least
one CAL deposit with energy \mbox{$E>400\MeV$} at pseudorapidity \mbox{$\eta_{max}>-1.5$}.

\section{Diffractive \mbox{$\bf M_X$} spectrum in rapidity--gap events}
\label{s:rap_gap}
\begin{figure}[tb]
\centerline{\hbox{
\psfig{figure=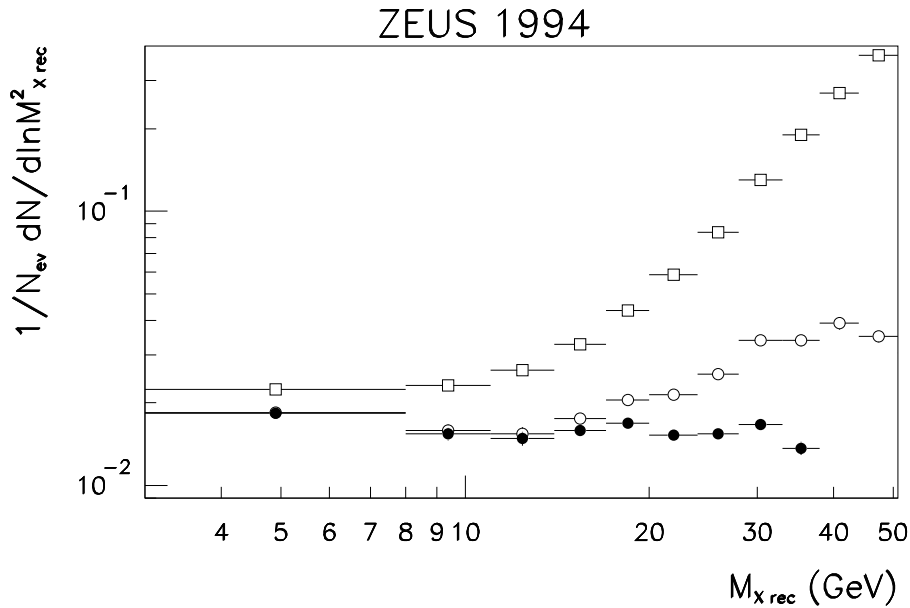}}}
\bf\caption
{\it 
Uncorrected spectrum of reconstructed hadronic mass in
photoproduction events at \mbox{$W\approx 200\GeV$} before (squares) and
after (open circles) imposing the requirement of no hits in the PRT.
The latter spectrum was subject to subtraction of the nondiffractive
contamination resulting in the uncorrected distribution attributed
to diffractive processes (solid points).
}
\label{f:data0}
\end{figure}

The spectrum of the reconstructed hadronic mass is shown in
figure~\ref{f:data0} (open squares).  It is presented in the form:
\begin{equation}
\frac{1}{N_{ev}}\cdot \frac{\Delta N}{\Delta \ln M^2_{X\:rec}},
\label{e:mx_raw}
\end{equation}
where \mbox{$\Delta N$} denotes the number of events 
reconstructed in a given \mbox{$\Delta \ln M^2_{X\:rec}$} interval
and \mbox{$N_{ev}$} is the total number of events accepted by
the trigger and passing the general
selection criteria described in sec.~\ref{s:ev_sel}.  
The variable bin width, \mbox{$\Delta \ln M^2_{X\:rec}$}, 
was adjusted such as to keep the purity for diffractive events above 
\mbox{$70\%$}. 

Diffractive processes are expected to give a contribution that is
approximately flat in such a double logarithmic plot,
and should dominate the region of low masses.
The steep rise of the spectrum at higher values of \mbox{$M^2_{X\:rec}$} is
due to a large contribution from nondiffractive processes.

\subsection{Selection of rapidity--gap events}
\label{ss:prt-rap-gap}
In order to suppress the contamination from nondiffractive processes, 
only the events with a forward gap in the rapidity distribution 
of the final state hadrons were used.
This rapidity--gap cut rejected
all events with hits in the PRT detector. A hit
was defined as a coincidence of signals of at least \mbox{$50$}~ADC counts
from both scintillator counter layers.  
The \mbox{$50\counts$} threshold should be
compared to the r.m.s. of the apparatus noise of \mbox{$17\counts$} and to 
the \mbox{$70-100\counts$} corresponding to a minimum ionizing particle.
The uncorrected \mbox{$M_{X\: rec}$} distribution in
events with no hits in the PRT is also shown in figure~\ref{f:data0} as the 
open circles.  
The PRT cut softens the rise at high values of \mbox{$M_{X\: rec}$}
by reducing the contribution from nondiffractive processes.  The 
remaining rise of the open circle points comes from nondiffractive processes
which do not produce hits in the PRT counters.  Before we correct for those, we
first describe the response of the PRT detector to nondiffractive
photoproduction processes.

\subsection{PRT response to nondiffractive processes}

The efficiency of the PRT counters to veto nondiffractive events was
studied with the EPSOFT and PYTHIA MC generators. 
It was found that there are two
factors which affect this efficiency.  The first factor is related to the
correlation between the multiplicity of particles produced in the
PRT angular region (\mbox{$4.3 < \eta < 5.8$}) and the invariant mass
of hadrons emitted in the angular region covered by the CAL.
This multiplicity diminishes with decreasing mass of the hadronic system.
Such behaviour is partially due to nondiffractive processes
with pion and reggeon exchange that contribute mainly to the region of
low \mbox{$M_{X\:rec}$} and may produce events with rapidity--gaps.
As a consequence, the
fraction of nondiffractive events which have a particle (with energy
above 1 GeV) emitted into the angular region of the PRT decreases
from 99\% at very high masses 
(\mbox{$M_{X\: rec}\sim 70\GeV$}), 
to about 85\% at intermediate masses (\mbox{$\sim 20\GeV$}) and to
\mbox{$\sim$} 75\% at lower masses (\mbox{$\sim 12\GeV$}).
 
The second factor which affects the efficiency comes from the 
particle absorption in the material in front of the PRT. Using a
detailed modelling of the detector in the beam pipe region, the probability 
for a particle emitted inside the PRT acceptance 
to give a coincidence signal in the two layers of the detector
was determined as a function of the
particle production angle and energy. It was found that particles with
\mbox{$5.0 < \eta < 5.8$} have a probability of more than 99\% to produce a
coincidence signal in the PRT.  Particles with \mbox{$4.3 < \eta < 5.0$} have a high
probability to be absorbed before reaching the PRT and on average only
30\% of them will produce a coincidence signal.
    
\begin{figure}[tb]
\centerline{\hbox{
\psfig{figure=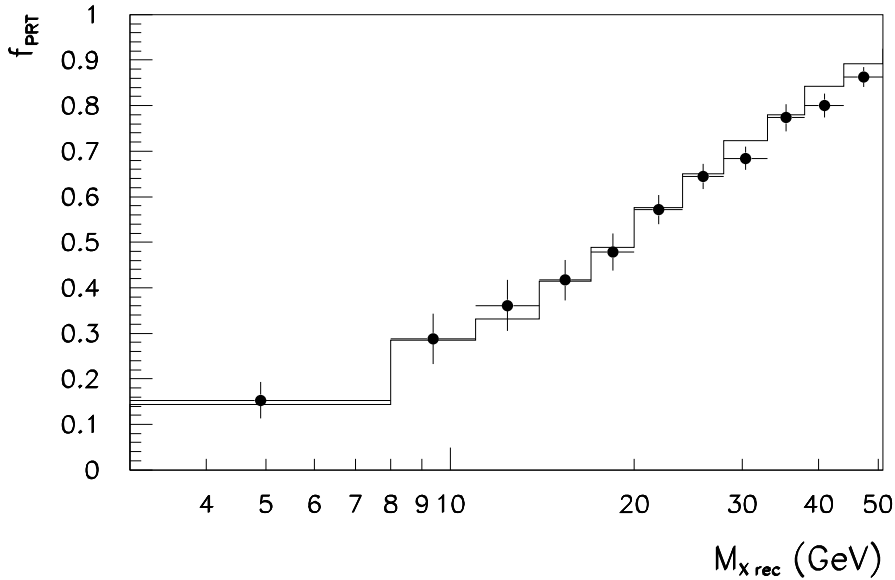}}}
\bf\caption{\it 
  The fraction of events with a PRT hit as function of the
  reconstructed mass \mbox{$M_{X\:rec}$} for data (solid points) and for MC
  (line).  
}
\label{fig:comp}
\end{figure}

The efficiency of the PRT to detect nondiffractive events depends on
the two effects described above. In the MC simulation
this efficiency varies from 95\% at very high masses to about 75\%
at \mbox{$M_{X\: rec}\sim 20\GeV$} 
and 65\% at \mbox{$M_{X\: rec}\sim 12\GeV$}.  
The reliability of the MC simulation
was tested by comparing the fraction of events with the PRT tag as a
function of the invariant mass observed in the CAL for the data and the
MC simulation, including all the photoproduction subprocesses. 
This comparison is presented in figure~\ref{fig:comp} and shows
good agreement between data and MC over the whole mass region used in
this study.

The sensitivity of the results to the noise and the inefficiencies of
the counters was investigated by repeating the whole analysis using
slightly modified criteria for rejecting events with particle activity
in the PRT: no coincidence between the two scintillator layers was
required and the events with more than \mbox{$50\counts$} signal in either of
the counters were rejected.  The difference between the results
obtained using this and the original selection method was used for the
estimate of the systematic uncertainty.

\subsection{Subtraction of remaining nondiffractive contribution}
\label{s:sub_nd}
The contribution of nondiffractive processes that survived the
PRT rapidity--gap cut was estimated by using a MC simulation technique.
By using the nondiffractive and 
the diffractive MC samples combined according to the procedure
described in sec.~\ref{ss:mc_comb}, the fraction of the
cross section due to diffractive reactions was calculated
for each \mbox{$\ln M_{X\: rec}^2$} bin.  This was then used to scale the 
measured spectrum, resulting in the distribution 
shown in figure \ref{f:data0} as solid points.  This
distribution needs to be corrected for acceptance, as discussed below.
Note that the subtraction is reliable at low masses where the nondiffractive
component is small.  As the mass grows the nondiffractive
contribution increases, reaching \mbox{$40\%$} of the signal
at \mbox{$M_{X\:rec}\approx 24\GeV$} 
making the measurements beyond this point quite model dependent.

\subsection{Acceptance for diffractive processes} 
\begin{figure}[tb]
\centerline{\hbox{
\psfig{figure=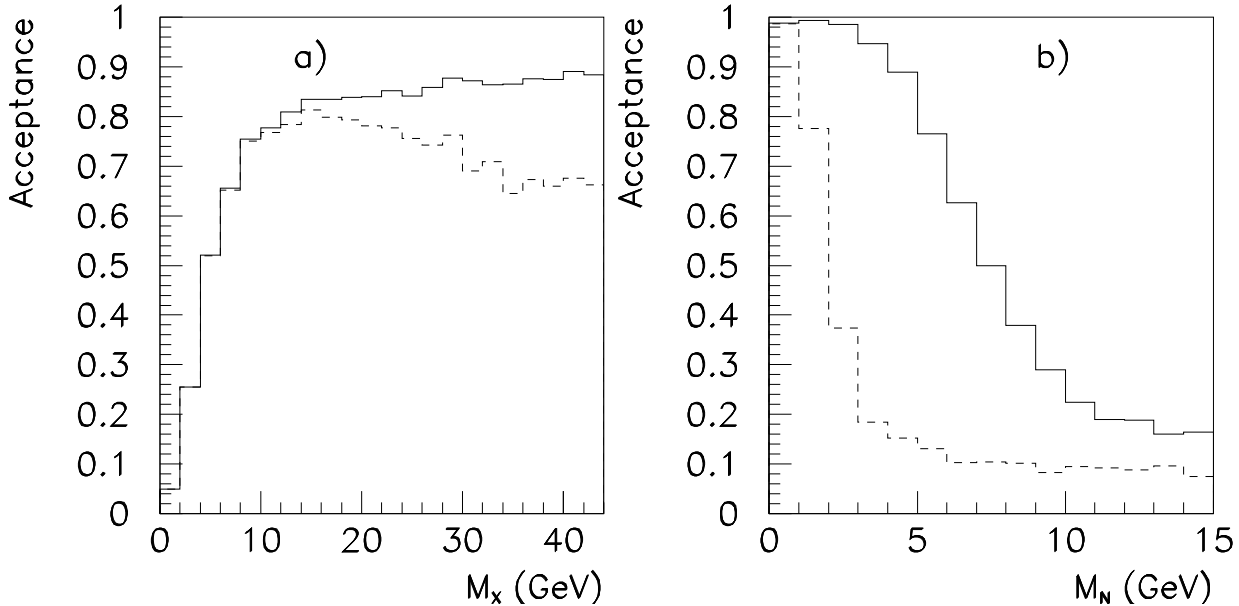}}}
\bf\caption{\it 
\mbox{$a$}) The combined acceptance of
the trigger and the \mbox{$\eta_{max}>-1.5$} cut for diffractive
processes obtained from MC simulation (solid line).  
The effect of adding the
requirement of no hits in PRT is also shown (dashed line). 
\mbox{$b$}) The acceptance for proton dissociation,
calculated as the fraction of events due to the process
\mbox{$\gamma p \rightarrow X N$} that are reconstructed with low mass in CAL, 
\mbox{$M_{X\:rec}<24\GeV$}, (solid line) and have no hits in the PRT
(dashed line).
}
\label{f:mx_acc}
\end{figure}

The combined acceptance of
the calorimeter trigger and the selection cuts for diffractive 
\mbox{$\gamma p\rightarrow Xp$}
events with \mbox{$176<W<225\GeV$} and \mbox{$Q^{2}< 0.02 \GeV^{2}$}
is presented in figure \ref{f:mx_acc}a as a
function of \mbox{$M_X$}.  
At low masses, in particular in the region
of light vector meson production, the acceptance is very low due to CAL trigger
inefficiency and the
\mbox{$\eta_{max}>-1.5$} cut.  For \mbox{$M_X>10\GeV$} the acceptance rises to
over \mbox{$80\%$}, where the trigger inefficiency is the main limiting factor.
If the rapidity--gap cut based on the PRT is imposed, the
acceptance for the diffractive photon dissociation in 
the mass region used for the measurement, 
\mbox{$M_X<24\GeV$},
changes by less than \mbox{$7\%$}.  For larger \mbox{$M_X$} the acceptance 
falls since the 
particles from the decay of the dissociated photon system reach the PRT.

In figure \ref{f:mx_acc}b the acceptance for the proton dissociation events
\mbox{$\gamma p\rightarrow XN$} with \mbox{$M_X<24\GeV$} that appear in the sample
of events with reconstructed mass \mbox{$M_{X\:rec}<24\GeV$} is shown.  
With increasing \mbox{$M_N$} the acceptance diminishes, falling below \mbox{$50\%$} for
\mbox{$M_N>7\GeV$}, since the particles from the decay
of the system \mbox{$N$} reach the CAL and thus \mbox{$M_{X\:rec}$}
is artificially large and beyond the region under study.
If the PRT cut is used, the acceptance drops below \mbox{$50\%$}
already at \mbox{$M_N\approx 2\GeV$}, resulting in a lower contribution 
of proton dissociation processes.

\subsection{Acceptance correction}

The uncorrected diffractive mass spectrum shown as solid points
in figure~\ref{f:data0} was corrected for
detector effects by means of a multiplicative correction function,
calculated using the MC:
\begin{equation}
Corr(M_X) = 
\left(\frac{1}{N_{gen}}\cdot 
      \frac{\Delta N^{diff}_{gen}}{\Delta \ln M^2_{X\:gen}}\right) /
\left(\frac{1}{N_{rec}}\cdot 
      \frac{\Delta N^{diff}_{rec}}{\Delta \ln M^2_{X\:rec}}\right).
\label{e:mx_correction}
\end{equation}
Here \mbox{$\Delta N^{diff}_{rec}$} denotes the number of diffractive 
MC events with no hits in the PRT
that are reconstructed in the considered 
\mbox{$\Delta \ln M^2_{X\:rec}$} interval
and \mbox{$N_{rec}$} is the number of MC events 
(including nondiffractive processes) passing 
the trigger and the general selection criteria.
The quantity \mbox{$N_{gen}$} denotes the total number of MC events
used for the calculation (including nondiffractive processes), 
while \mbox{$\Delta N^{diff}_{gen}$} is the number of diffractive 
MC events with \mbox{$M_N<2\GeV$} 
that were generated in the interval \mbox{$\Delta \ln M^2_{X\:gen}$}.
The value of the correction factor \mbox{$Corr(M_X)$} is in the range
\mbox{$0.9-1.1$} apart from the first mass bin, \mbox{$3<M_{X\:rec}<8\GeV$},
where it is close to \mbox{$1.6$}.
This method corrects for the following effects:
\begin{itemize}
\item
the limited trigger acceptance and the inefficiencies of the event 
selection cuts (see solid line in figure~\ref{f:mx_acc}a);
\item
the reduction of the acceptance for diffractive processes in the region
of high \mbox{$M_X$} values due to PRT cut (see dashed line 
in figure~\ref{f:mx_acc}a);
\item
smearing of the limit on the nucleonic mass \mbox{$M_N<2\GeV$} due to the
PRT acceptance (see figure~\ref{f:mx_acc}b);
\item
migration effects in the mass reconstruction procedure.
\end{itemize}

\subsection{Corrected \mbox{$\bf M_X$} spectrum in diffractive events}
\begin{figure}[tb]
\centerline{\hbox{
\psfig{figure=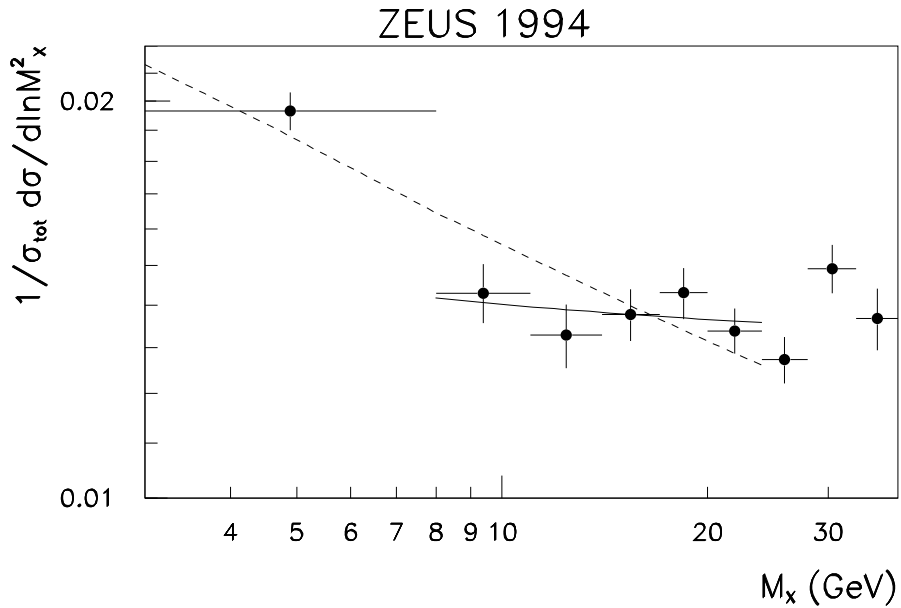}}}
\bf\caption[Corrected rapidity--gap \mbox{$M_X$} spectrum]
{\it
The acceptance corrected \mbox{$M_X$}
spectrum in diffractive photoproduction events.
The combined statistical errors of the
data and of the correction factors are shown as vertical bars.
The line indicates the result of the fits of Eq.~(\protect\ref{e:mx-for-fit})
to the diffractive spectrum in the intervals 
\mbox{$3<M_X<24\GeV$} (dashed line) and \mbox{$8<M_X<24\GeV$} (solid line).
}
\label{f:corr_data2}
\end{figure}
Figure~\ref{f:corr_data2} presents the corrected \mbox{$M_X$} spectrum in
diffractive photoproduction events.
The quantity plotted corresponds to the probability per unit 
\mbox{$\ln M^2_X$} that a
photoproduction event with \mbox{$W\approx 200\GeV$} is due to
a diffractive process \mbox{$\gamma p\rightarrow XN$}, where \mbox{$N$} is 
either a proton or a nucleonic system with \mbox{$M_N<2\GeV$}.
Apart from the first bin, \mbox{$3<M_X<8\GeV$}, the distribution is
flat in the double logarithmic plot as
expected for the diffractive processes dominated by the 
triple pomeron amplitude.

\section{Analysis of \mbox{$\bf M_X$} spectrum in rapidity--gap events}
\label{s:anal_rap_gap}
\subsection{Diffractive cross section}

By summing over the contents of the bins of the 
mass spectrum of figure \ref{f:corr_data2} 
the fraction of the total
photoproduction cross section attributed to
photon dissociation, \mbox{$\gamma p \rightarrow XN$}, with
\mbox{$3<M_X<24\GeV$} and \mbox{$M_N<2\GeV$} is determined to be
\mbox{$ \sigma^{partial}_{D}/\sigma_{tot} = 6.2 \pm 0.2 (stat) \%.$}

\subsection{Shape of the \mbox{$M_X$} spectrum}
The diffractive mass spectrum obtained from the rapidity--gap data
was fitted in the range \mbox{$3<M_X<24\GeV$} with 
Eq.~(\ref{e:tripple-Regge}) evaluated 
for the triple pomeron case, \mbox{$ijk=\po\po\po$}, and integrated over \mbox{$t$} up to the kinematic limit \mbox{$t_{max}$}:
\begin{equation}
\frac{d \sigma}{d \ln M_{X}^2} = 
M_{X}^2 \frac{d \sigma}{d M_{X}^2} = 
M_{X}^2 \int_{-\infty}^{t_{max}}  \frac{d^2 \sigma}{d t d M_{X}^2} d t 
\propto 
\frac{M_{X}^2}{b_{\circ} + 2 \alphapomp \ln \frac{W^2}{M_{X}^2}} \cdot
\left(\frac{1}{M_{X}^2}\right) ^{\alphapom(0)}.
\label{e:mx-for-fit}
\end{equation}
The parameter values of \mbox{$\alphapomp=0.25\GeV^{-2}$}~\cite{Goulianos}
and \mbox{$b_{\circ}=4\GeV^{-2}$}~\cite{Chapin} were assumed in accord with
results of experiments at lower energies.
For the fit the function was integrated over
each of the bins and the obtained values were compared with the corresponding
number of data events.   

As the result of the fit
a pomeron intercept of \mbox{$\alphapom(0)=1.20\pm 0.02 (stat)$} was obtained, 
although with a poor \mbox{$\chi^2$} (see dashed curve in figure \ref{f:corr_data2}).
A similar fit performed only for the range \mbox{$8<M_X<24\GeV$}
gives a lower value of \mbox{$\alphapom(0)=1.12\pm 0.04 (stat)$} and
provides a good description of the data in the fitted mass interval
(see solid curve in figure \ref{f:corr_data2}).
As the values of the pomeron intercept obtained from the triple--pomeron
fit show some dependence on the fitting interval, this
parameter will be referred to as an effective intercept.

\subsection{The \mbox{$\po\po\re$} component}
\label{ss:ppr}
\begin{figure}[tb]
\centerline{\hbox{
\psfig{figure=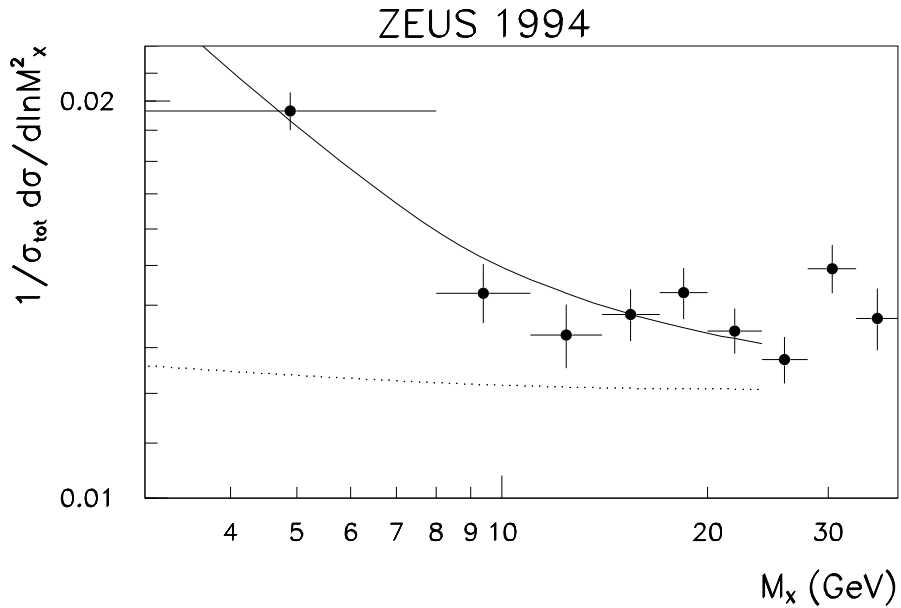}}}
\bf\caption
{\it
The acceptance corrected \mbox{$M_X$} spectrum in
diffractive photoproduction events fitted with the sum of a
\mbox{$\po\po\po$} and a \mbox{$\po\po\re$} triple Regge terms (solid line).
The contribution of the triple pomeron term is shown as a dotted line.
}
\label{f:corr_data3}
\end{figure}

If the function fitted in the range \mbox{$8<M_X<24\GeV$} is extrapolated
to the region of the first bin, it falls significantly below the data point.
A possible explanation 
is the contribution of a \mbox{$\po\po\re$} term in addition
to the triple pomeron amplitude.  The precision of the data is
insufficient to perform a reliable fit to the sum of the 
two components and to determine their relative contributions,
as well as the pomeron and the reggeon intercepts.
We have nevertheless verified that the obtained spectrum
is consistent with the intercepts \mbox{$\alphapom(0)=1.08$} and 
\mbox{$\alphareg(0)=0.45$} derived from fits to total 
and elastic cross sections.
Assuming these values we performed
a fit of the sum of the \mbox{$\po\po\po$} and the 
\mbox{$\po\po\re$} terms to the entire interval of \mbox{$3<M_{X}<24\GeV$}.
The fit indicates that the \mbox{$\po\po\re$} term amounts to
\mbox{$26\pm 3(stat)\%$} 
of the diffractive cross section in the considered \mbox{$M_X$} range.
As shown in figure \ref{f:corr_data3} 
the function obtained (solid line) is in good agreement with the data.
The dotted line shows the fraction of the cross section attributed to
the triple pomeron term alone, with an assumed pomeron intercept of
\mbox{$\alphapom(0)=1.08$}.

\subsection{Single diffractive cross section} 
As discussed earlier, we measured the relative cross section 
for the diffractive process
\mbox{$\gamma p \rightarrow X N$} with \mbox{$3<M_X<24\GeV$} and \mbox{$M_N<2\GeV$} 
as well as the shape of the \mbox{$M_X$} spectrum in this kinematic region.
Using these data we estimated the fraction of the total
photoproduction cross section that can be attributed to
single diffractive photon dissociation.
This was done by means of an analytic extrapolation
of the \mbox{$M_X$} spectrum beyond the measured interval
using the parameterization based on the sum of the \mbox{$\po\po\po$}
and the \mbox{$\po\po\re$} terms.
The parameterization was integrated from the \mbox{$\phi$} meson mass, the
heaviest of the three light vector mesons contributing to 
elastic photoproduction, up to \mbox{$M_X^2=0.05\cdot W^2$}.
The small contribution of proton excitation with \mbox{$M_N<2\GeV$} was corrected
for by assuming that the probability for the proton to be excited
is the same as in \mbox{$pp$} reactions, namely \mbox{$5-6\%$}~\cite{Goulianos}.
The resulting ratio of the cross section for the single diffractive 
photon dissociation to the total photoproduction cross section is
\mbox{$\sigma_{SD}/\sigma_{tot} = 13.3 \pm 0.5 (stat)\%.$}

\subsection{Systematic uncertainties}
\label{ss:rg-sys}
The systematic uncertainties in the results were studied by repeating
the analysis using various event selection and acceptance correction
methods and by changing the fit parameters within reasonable limits.
The difference in the obtained results was 
used as an estimate of the uncertainty. 
Table~\ref{t:mx1_fit_results1} summarizes the outcome of the
checks.
\begin{table}[tb]
\centerline{\hbox{
\begin{tabular}{|c||c|c|c|}
\hline
source of the uncertainty &
\mbox{$|\Delta \sigma^{partial}_{D}/\sigma_{tot}|$} &
\mbox{$|\Delta \alphapom(0)|$} & 
\mbox{$|\Delta f_{\pom\pom\reg}|$} \\
\hline\hline
nondiffractive contamination           & \mbox{$0.9\%$} & \mbox{$0.06$} & 9\% \\
PRT noise and efficiency                 & \mbox{$0.5\%$} & \mbox{$0.05$} & 7\% \\
diffractive MC model                     & \mbox{$0.7\%$} & \mbox{$0.03$} & 1\%\\
vector meson production cross section    & \mbox{$0.1\%$} & \mbox{$0.01$} & 2\%\\
\mbox{$M_{X}$} reconstruction                   & \mbox{$0.1\%$} & \mbox{$0.01$} & 1\% \\
increased CAL FLT thresholds             & \mbox{$0.1\%$} & \mbox{$0.01$} & 1\%\\
\mbox{$M_X$} fit interval                       &  & \mbox{$0.01$} & 2\%\\
value of \mbox{$b_{\circ}$}                     &  & \mbox{$0.01$} & 2\%\\
value of \mbox{$\alphapomp$}                   &  & \mbox{$0.01$} & 1\%\\
\hline
\end{tabular}
}}
\bf\caption
{\it 
Individual contributions to the systematic uncertainty in the results.
}
\label{t:mx1_fit_results1}
\end{table}

The largest uncertainty is due to the dependence of the result
on the way the nondiffractive contribution is modelled.
This effect was estimated by using PYTHIA to simulate the nondiffractive 
interactions instead of EPSOFT (see sec.~\ref{s:mc}).
Another important source of systematic error is related 
to the noise and the efficiency of the PRT.
The corresponding uncertainty was evaluated by repeating
the analysis without requiring the coincidence between the two
layers of counters, as discussed in section~\ref{s:ev_sel}.
The sensitivity to the model of diffractive processes used for the 
acceptance correction was checked by repeating the analysis using
the EPSOFT generator instead of the NZ one (see sec.~\ref{s:mc}).

In addition to these dominant sources of systematic error
a number of other effects were studied.
The cross section for vector meson production was 
changed by the size of the error, i.e. \mbox{$\pm 3\%$}, to check how the
results depend on the simulated
\mbox{$M_X$} behaviour in the region of low mass resonances (see sec.~\ref{s:mc}).
The sensitivity to the precision of the diffractive mass reconstruction 
was verified by using an alternative method of \mbox{$M_{X}$} determination
that does not rely on low energy CAL condensates (see sec.~\ref{s:mx_rec}).
To estimate the uncertainty due to
imprecise calibration of the CAL trigger, the whole analysis 
was repeated with higher trigger thresholds.
The energy thresholds applied to the data and to the MC 
were raised from \mbox{$464\MeV(1250\MeV)$} to \mbox{$660\MeV(1875\MeV)$}
for the RCAL EMC trigger excluding (or including) the towers adjacent to
the beam pipe.
To check the stability of the fitting procedure,
the fit was repeated using mass intervals extended to \mbox{$M_{X}=32\GeV$}.
To examine the dependence of the result on the assumed parameter values,
the \mbox{$t$} slope parameter was changed from \mbox{$b_{\circ}=4\GeV^{-2}$}
to \mbox{$5\GeV^{-2}$}
and \mbox{$\alphapomp=0.20\GeV^{-2}$} was used instead of the
original value of \mbox{$0.25\GeV^{-2}$}.
To estimate the overall systematic uncertainty,
all contributions  were added in quadrature.

\subsection{Results}
\label{ss:rg-results}
We summarize here the results of the analysis based on the rapidity--gap data
in photoproduction at \mbox{$W\approx 200\GeV$}:
\begin{itemize}
\item
The fraction of the total photoproduction cross section attributed to
the photon dissociation, \mbox{$\gamma p\to XN$}, 
in the mass ranges \mbox{$3<M_X<24\GeV$} and \mbox{$M_N<2\GeV$} is:
\[\sigma^{partial}_{D}/\sigma_{tot} = 6.2 \pm 0.2 (stat) 
   \pm 1.4 (syst)\%.\]
\item
The effective pomeron intercept derived by 
fitting the diffractive mass spectrum in the range \mbox{$8<M_X<24\GeV$} with 
the triple pomeron expression is:
\[\alphapom(0)=1.12 \pm 0.04(stat) \pm 0.08(syst).\]
\item
If the data in the region \mbox{$3<M_X<24\GeV$} are fitted by the sum of two
pomeron exchange terms, \mbox{$\po\po\po$} and \mbox{$\po\po\re$}, assuming
\mbox{$\alphapom(0)=1.08$} and \mbox{$\alphareg(0)=0.45$}, the fraction of the
\mbox{$\po\po\re$} term with respect to the sum of the two terms is:
\[f_{\pom\pom\reg}=26\pm 3(stat) \pm 12(syst)\%.\]
\item
The fraction of the total photoproduction cross section due to single 
diffractive photon dissociation, \mbox{$\gamma p\to Xp$}, in the mass range
\mbox{$\m_{\phi}^2<M_X^2<0.05W^2$} is:
\[\sigma_{SD}/\sigma_{tot}=13.3 \pm 0.5 (stat) \pm 3.6 (syst)\%.\]
\end{itemize}

\section{ Analysis of the inclusive \mbox{$\bf M_X$} spectrum}
\label{s:lnMx2_anal}
In the analysis described above the identification of the
diffractive processes was
based on the rapidity--gap signature.  The remaining
contamination from nondiffractive events satisfying the
rapidity--gap requirement was corrected for by using
the MC simulation.  To test the sensitivity of the results
to the model assumptions made in that study
we performed another analysis of the same data sample, with 
a different approach.  This time no
rapidity--gap was required and the distinction between the different
processes was performed solely on the basis of the shape of the mass
spectrum~\cite{ZEUS-DIS-diff}.  This alternative study is described below.

\subsection{ Uncorrected spectrum}
\begin{figure}[tb]
  \centerline{\hbox{ \psfig{figure=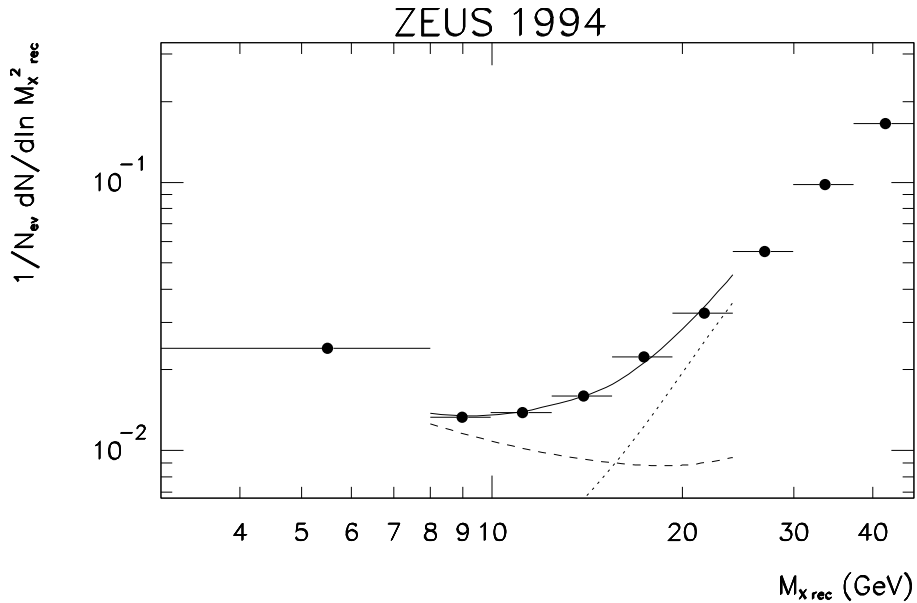}}} 
\bf\caption {\it
    The uncorrected distribution of reconstructed hadronic mass in
    photoproduction events at \mbox{$W\approx 200\GeV$}.  The solid curve
    shows the result of the fit with the sum of the diffractive and the
    nondiffractive components, which are also shown separately as
    dashed and dotted curves, respectively.  The curves correspond to
    parameterizations (\protect{\ref{e:lnmx_nd}}) and
      (\protect{\ref{e:lnmx_d}}) folded with the detector acceptance correction
    factors.  }
\label{f:lnmx2} 
\end{figure}

This analysis was performed on the same
sample of photoproduction events used in the previous analysis
and selected according to the criteria described in sec.~\ref{s:ev_sel}.
However,
neither the rapidity--gap requirement
nor the \mbox{$\eta_{max}>-1.5$} cut were imposed.
This data sample was used to determine the distribution of the
reconstructed hadronic mass (see sec.~\ref{s:mx_rec} for details of
the mass reconstruction):
\begin{equation}
\frac{1}{N_{ev}} \frac{\Delta N}{\Delta \ln M^2_{X\:rec}},
\label{e:ln_mxrec}
\end{equation}
where \mbox{$N_{ev}$} is the total number of selected events and 
\mbox{$\Delta N$} denotes the number of events reconstructed in the
given \mbox{$\ln M^2_{X\:rec}$} bin.  The size of the bins \mbox{$\Delta \ln M^2_{X\:rec}$}
was chosen in a similar way to that used in the previous analysis in
order to
limit the bin--to--bin migrations.  The reconstructed
mass spectrum obtained in this way is presented in figure \ref{f:lnmx2}.

\subsection{ Determination of diffractive and nondiffractive components}
The uncorrected mass spectrum was fitted 
with the sum of the diffractive and nondiffractive components:
\begin{equation}
\frac{1}{N_{ev}} \frac{\Delta N}{\Delta \ln M^2_{Xrec}} =\\
A_{ND} \cdot \frac{1}{\sigma_{tot}} 
        \frac{\Delta {\sigma}_{ND}}{\Delta \ln M^2_X} +
A_{D} \cdot \frac{1}{\sigma_{tot}} 
        \frac{\Delta {\sigma}_{D}}{\Delta \ln M^2_X},
\label{e:ln_mxrec_sum}
\end{equation}
where \mbox{$A_{ND}$} and \mbox{$A_{D}$} are \mbox{$M_X$} dependent
correction factors 
for the nondiffractive and the diffractive components,
respectively.  They account for the limited
acceptance of the trigger and selection cuts as well as for the
effects of migrations between the true and the reconstructed mass bins.
The true value of \mbox{$M^2_X$} for the case of nondiffractive processes
was defined as the total invariant mass of hadrons emitted
in the angular region covered by the CAL which corresponds to
\mbox{$-3.8<\eta<4.3$}.  The correction factors were obtained from
the MC simulation using EPSOFT
for the case of nondiffractive processes,
and NZ for the diffractive processes.

Similarly to the approach presented in \cite{ZEUS-DIS-diff}, 
we have assumed that
the nondiffractive contribution in the region 
\mbox{$M^2_X<$} 0.05 \mbox{$ W^2$} may be parameterized by a single
exponential form:
\begin{equation}
\frac{1}{\sigma_{tot}}\frac{d{\sigma}_{ND}}{d \ln M^2_X}=
C \exp (B \ln M^2_X).
\label{e:lnmx_nd}
\end{equation}
This form can be understood from the assumption of uniform, uncorrelated
particle emission in rapidity space. 
The corresponding \mbox{$M_X$} distribution is
directly related to fluctuations in the number of
particles emitted in the angular region covered by CAL.
The slope \mbox{$B$} and the normalization factor \mbox{$C$} were determined by the fit.

For the diffractive component in Eq.~(\ref{e:ln_mxrec_sum})
a triple pomeron component integrated over \mbox{$t$} 
was assumed (see also Eq.~(\ref{e:mx-for-fit})):
\begin{equation}
\frac{1}{\sigma_{tot}}\frac{d{\sigma}_{D}}{d\ln M^2_X}=
D 
\frac{1}{b_{\circ} + 2 \alphapomp \ln \frac{W^2}{M_{X}^2}} 
\left(\frac{1}{M_{X}^2}\right) ^{\alphapom(0)-1}.
\label{e:lnmx_d}
\end{equation}
The  same values of the parameters  \mbox{$b_{\circ}$} and \mbox{$\alphapomp$} as in the
rapidity--gap analysis were assumed.  The pomeron intercept, \mbox{$\alphapom(0)$},
defining the slope of the diffractive mass distribution and the 
normalization factor \mbox{$D$} were determined from the fit.

The result of the fit of expression (\ref{e:ln_mxrec_sum}) 
to the uncorrected mass spectrum in the range \mbox{$8<M_{X\:rec}<24\GeV$}
is presented in figure \ref{f:lnmx2} as a solid curve.
The dashed and dotted curves represent the 
diffractive and the nondiffractive contributions, respectively.

The slope and the magnitude of the diffractive component
is  constrained by the low mass behaviour of the measured spectrum.
The fit gave a value of 
\mbox{$\alphapom(0) = 1.15\pm 0.08 (stat)$}.
The parameters of the nondiffractive term (Eq.~(\ref{e:lnmx_nd}))
are driven by the rise of the 
spectrum at large masses where this component dominates.
The fit result for the value of the nondiffractive slope
is \mbox{$B= 1.30 \pm 0.08 (stat)$}.

In the analysis of the rapidity--gap data we observed that in order
to describe the diffractive mass spectrum including the low mass
region, \mbox{$3<M_X<8\GeV$}, a significantly higher value of the effective
pomeron intercept
is needed.  To verify this observation we extended the mass interval used 
in the present analysis to include also this region.
The value obtained for the effective
pomeron intercept was \mbox{$\alphapom(0) = 1.25 \pm 0.08 (stat)$}, significantly
higher than the one obtained previously, though the fit resulted in a
poor \mbox{$\chi^2$}.
As suggested already in sec.~\ref{ss:ppr} this 
discrepancy may be due to a contribution of another diffractive term,
the \mbox{$\po\po\re$}, in addition to the triple pomeron component.
To further test this hypothesis we repeated the
fit of the sum of the diffractive and the nondiffractive components
to the uncorrected mass spectrum assuming that the
diffractive part is a sum of the \mbox{$\po\po\po$} and the \mbox{$\po\po\re$} terms.
As in the first analysis, the precision of the data
was insufficient to determine the individual subprocess contributions
as well as the pomeron and reggeon intercepts.
Therefore, we assumed the values of
\mbox{$\alphareg(0)=0.45$} and \mbox{$\alphapom(0)=1.08$} and determined the
fraction \mbox{$f_{\pom\pom\reg}$} of the diffractive cross section in the mass
interval used for the fit,
\mbox{$3<M_X<24\GeV$}, due to the  \mbox{$\po\po\re$} term.  This
was found to be \mbox{$f_{\pom\pom\reg}= 23 \pm 5 (stat)\%$}.

Based on the results of the latter fit we derived the 
fraction of the total photoproduction cross section attributed to
the diffractive process, \mbox{$\gamma p \rightarrow X N$}, 
where  \mbox{$3<M_X<24\GeV$} and \mbox{$M_N<5\GeV$}, to be
\mbox{$\sigma^{partial}_{D}/\sigma_{tot} = 5.0 \pm 0.2 (stat)$}.
The limit on the mass of the dissociated proton state is higher than
in the case of the rapidity--gap analysis.  This is due to the
absence of the PRT veto which rejects a large fraction of
events with higher \mbox{$M_N$}, as
may be seen from the acceptance plot in figure \ref{f:mx_acc}.
If the diffractive component obtained from the fit is 
integrated over the mass range
\mbox{$m^2_\phi<M_X^2<0.05 W^2$}, the cross section for single diffractive photon
dissociation relative to the total photoproduction cross section
is found to be \mbox{$\sigma_{SD}/\sigma_{tot} = 11.0 \pm 0.5 (stat)\%$}.

\subsection{Systematic checks} 

The analysis of systematic uncertainties carried out for these results
was similar to that of the rapidity--gap
analysis already described in sec.~\ref{ss:rg-sys}.
Here, we concentrate only on the two elements that were different.

As Eqs.~(\ref{e:lnmx_nd}) and (\ref{e:lnmx_d}) are 
expected to describe the data up to
\mbox{$M^2_X<$} 0.05 \mbox{$\cdot W^2$}, we repeated the analysis moving the upper
limit on the fitting interval from \mbox{$24\GeV$} to \mbox{$40\GeV$}.
No significant change in the results of the fit was observed.

We also studied the dependence of the results on the \mbox{$M_X$} binning 
used.  Introducing equal bins in \mbox{$M_X$} instead of
\mbox{$\ln M_X^2$} we did not observe significant changes in the fit results 
apart from the value of \mbox{$\alphapom(0)$} which moved by \mbox{$\pm 0.09$}.
This was included into the systematic error.

To estimate the overall systematic uncertainty,
all error contributions  were added in quadrature.

\subsection{Results and comparison with rapidity--gap analysis}

The results of this analysis
of photoproduction at \mbox{$W\approx 200\GeV$} are:
\begin{itemize}
\item 
The
fraction of the total photoproduction cross section attributed to
the photon dissociation, \mbox{$\gamma p\to XN$},  
in the mass ranges \mbox{$3<M_X<24\GeV$} and \mbox{$M_N<5\GeV$} is:
\[\sigma^{partial}_{D}/\sigma_{tot} = 5.0 \pm 0.2 (stat) \pm 2.0 (syst)\%.\]
\item
The
effective pomeron intercept derived from the fit of the triple pomeron relation
to the diffractive mass spectrum in the range \mbox{$8<M_X<24\GeV$} is:
\[\alphapom(0)=1.15 \pm 0.08(stat) \pm 0.12(syst).\]
\item
If the data in the region \mbox{$3<M_X<24\GeV$} are fitted by the sum of two
pomeron exchange terms, \mbox{$\po\po\po$} and \mbox{$\po\po\re$}, assuming
\mbox{$\alphapom(0)=1.08$} and \mbox{$\alphareg(0)=0.45$}, the fraction of the
\mbox{$\po\po\re$} term with respect to the sum of the two terms is:
\[f_{\pom\pom\reg}=23\pm 5(stat) \pm 15(syst)\%.\]
\item
The fraction of the total photoproduction cross section due to single 
diffractive photon dissociation, \mbox{$\gamma p\to Xp$}, in the mass range
\mbox{$\m_{\phi}^2<M_X^2<0.05W^2$} is:
\[\sigma_{SD}/\sigma_{tot} = 11.0 \pm 0.5 (stat)  \pm 5 (syst)\%.\]

\end{itemize}

These numbers should be compared to the results of the first analysis
presented in sec.~\ref{ss:rg-results}.  The first of the
results, the fraction of the total cross section attributed to
photon dissociation, obtained in this analysis should be
larger than that obtained in the rapidity--gap study  by
roughly \mbox{$5\%$}
due to the different \mbox{$M_N$} limit.  However, the precision of the data
is too low to establish this difference.

The results of the two analyses are in good agreement.
As it was already pointed out, the two analyses differ drastically
in the way the diffractive 
and nondiffractive processes are distinguished:
\begin{itemize}
\item
In the first analysis the
diffractive processes were identified using the rapidity--gap signature
and the remaining
contamination from nondiffractive processes was corrected for using
the MC simulation.  
This method relies on the MC programs to
simulate the effect of the rapidity--gap cut on the nondiffractive processes.
However, as the nondiffractive contamination of the final \mbox{$M_X$} 
distribution is smaller than in the second method
the precision of the results is higher.
\item
In the second analysis no
rapidity--gap was required and the distinction between the different
processes was performed solely on the basis of the shape of the mass
spectrum.  This approach does not require MC simulation 
of rapidity--gaps in nondiffractive events.  However,
the larger nondiffractive contamination results in 
lower precision of the results.
\end{itemize}

\section{Discussion and conclusions}
\label{s:concl}
Using the ZEUS detector, we studied the diffractive process
\mbox{$\gamma p \rightarrow X N$}, where \mbox{$N$} is either a proton or a nucleonic
system with \mbox{$M_N<2\GeV$}, in photoproduction at high c.m. energy,
\mbox{$W \approx 200\GeV$}.
Relying on the rapidity--gap signature to identify the diffractive 
processes we measured the mass spectrum of dissociated photon states
in the range \mbox{$3<M_X<24\GeV$}.  The results were confirmed in an 
analysis where the distinction between the diffractive
and nondiffractive processes was based entirely on the
shape of the mass spectrum. 

We measured the fraction of the total
photoproduction cross section attributed to the diffractive process
\mbox{$\gamma p \rightarrow X N$} where \mbox{$3<M_X<24\GeV$} and 
\mbox{$M_N<2\GeV$} to be:
\[\sigma^{partial}_{D}/\sigma_{tot} = 6.2 \pm 0.2 (stat) 
   \pm 1.4 (syst)\%.\]
By extrapolating beyond the measured \mbox{$M_X$} interval and
correcting for the small contribution of proton dissociation we estimated the
cross section for single diffractive photon dissociation, 
\mbox{$\gamma p \to Xp$},
relative to the total photoproduction cross section to be:
\[\sigma_{SD}/\sigma_{tot} 
= 13.3 \pm 0.5 (stat)  \pm 3.6 (syst)\%.\]  
This value is consistent
with those obtained from other measurements of photoproduction reactions
at HERA~\cite{H1-sigtot,ZEUS-sigtot} 
and with the ones measured in diffractive proton dissociation in 
\mbox{$p\bar{p}$} interactions at c.m. 
energies of \mbox{$\sqrt{s}=546\GeV$} and \mbox{$1.8\TeV$} at the
Tevatron~\cite{CDF-mx,CDF-sigtot}.

The shape of the diffractive \mbox{$M_X$} distribution in the region of
\mbox{$8<M_{X}<24\GeV$} can be parameterized
by the triple pomeron formula with an effective pomeron intercept of: 
\[\alphapom(0)=1.12\pm 0.04(stat) \pm 0.08(syst).\]
This value is in good agreement with those obtained from the
shape of the diffractive mass spectrum measured
in \mbox{$p\bar{p}$} interactions at the Tevatron~\cite{CDF-mx}.  

The cross section for photon dissociation at low masses,
\mbox{$3<M_X<8\GeV$}, is
significantly higher than that expected from the triple pomeron amplitude 
when using the value of
\mbox{$\alphapom(0)$} derived at higher masses.
This behaviour of the \mbox{$M_X$} spectrum  may be
due to the contribution of another triple Regge term describing
pomeron exchange, namely \mbox{$\po\po\re$}.  
We verified that the measured spectrum is well described by the
sum of the two components with intercepts of \mbox{$\alphapom(0)=1.08$} and 
\mbox{$\alphareg(0)=0.45$}~\cite{DL-sigtot} 
derived from fits to total and elastic hadronic cross sections.
If these values are assumed, 
the fit in the interval of \mbox{$3<M_{X}<24\GeV$} 
indicates that the \mbox{$\po\po\re$} term is responsible for:
\[f_{\pom\pom\reg}=26\pm 3(stat) \pm 12(syst)\%\]
of the diffractive cross section in the considered \mbox{$M_X$} range.
This size of the \mbox{$\po\po\re$} 
contribution is similar to that obtained from the global fits to 
diffractive dissociation \mbox{$pp$} data at low energies~\cite{Field-Fox}.

To conclude, these studies of the diffractive mass spectrum indicate that
the dissociation of real photons at a c.m. energy of
\mbox{$W\approx 200\GeV$} is
similar to that of hadrons, as expected from the VDM,
and it is well described by Regge phenomenology.

\section*{Acknowledgments}
We thank the DESY Directorate for their strong support and
encouragement.  The remarkable achievements of the HERA machine group were
essential to collect the data used for the present analysis.


\begin{thebibliography}{99}
%
\bibitem{Sakurai}
J.J.Sakurai, Ann.~Phys.~11 (1960) 1.
%
\bibitem{Bauer}
T.H.Bauer et al., Rev.~Mod.~Phys.~50 (1978) 261.
%
\bibitem{DL-sigtot}
A.Donnachie and P.V.Landshoff, Nucl.~Phys.~B244 (1984) 322;\\
P.V.Landshoff, Nucl.~Phys.~B(proc.Suppl) 18C (1990) 211.
%
\bibitem{H1-sigtot}
H1 Collab., S.Aid et al., Z.Phys. C69 (1995) 27.
%
\bibitem{ZEUS-sigtot} 
ZEUS Collab., M.~Derrick et al., Z.~Phys.~C63~(1994)~391.
%
\bibitem{ZEUS-rho}
ZEUS Collab., M.~Derrick et al., Z.Phys C69 (1995) 39.
%
\bibitem{H1-rho}
H1 Collab., S.Aid et al., Nucl.Phys. B463 (1996) 3.
%
\bibitem{Mueller}
A.H.Mueller, Phys.~Rev. D2 (1970) 2963.
%
\bibitem{Goulianos}
K.Goulianos, Phys.Rep. 101 (1983) 169;\\
K.Goulianos, Nucl.~Phys.~B (Proc.~Suppl.) 12 (1990) 110.
%
\bibitem{CDF-mx}
CDF Collab., F.Abe et al., Phys.~Rev.~D 50 (1994) 5535.
%
\bibitem{Field-Fox}
R.D.Field and G.C.Fox, Nucl.~Phys.~B80 (1974) 367.
%
\bibitem{Chapin} 
T.J.Chapin et al., { Phys.~Rev.}~{ D31}~(1985)~17.
%
\bibitem{H1-mx}
H1 Collaboration, C.Adloff et al., DESY 97-009, submitted to Z.~Phys.
%
\bibitem{Collins}
D.B.Collins, ``An Introduction to Regge Theory and High Energy Physics'',
Univ.~Press (1977).
%
\bibitem{Collins-Martin}
D.B.Collins, A.D.Martin, ``Hadron Interactions'', Adam Hilger (1984).
%
\bibitem{status93}
The ZEUS Detector, Status Report, DESY(1993).
%
\bibitem{ZEUS-description}
ZEUS Collab., M.~Derrick et al., Phys. Lett. { B293} (1992) 465.
%
\bibitem{CAL}
M.~Derrick et al., Nucl. Instr. Meth. { A309} (1991) 77;\\         
A.~Andresen et al., Nucl. Instr. Meth. { A309} (1991) 101;\\    
A.~Bernstein et al., Nucl. Instr. Meth. { A336} (1993) 23.
%
\bibitem{LUMI}
D.~Kisielewska et al., DESY-HERA report 85-25 (1985);\\
J.~Andruszkow et al., DESY~92-066 (1992);\\
K.~Piotrzkowski, PhD Thesis, Cracow INP-Exp, 1993, DESY~F35D-93-06.
%
\bibitem{LPS}
M.~Derrick et al., DESY 96-183 (1996), submitted to Z.~Phys.
%
\bibitem{Burow}
B.D.Burow, PhD Thesis, University of Toronto, DESY F35D-94-01 (1994).
%
\bibitem{NZ}
N.N.~Nikolaev and B.G.~Zakharov, Z.~Phys.~{ C53}~(1992)~331.
%
\bibitem{Solano}
P.~Bruni et al., Proc. Workshop on Physics at HERA, DESY (1991)~363;\\
A.~Solano, PhD Thesis, University of Torino, 1993.
%
\bibitem{EPSOFT}
M.Kasprzak, PhD thesis, Warsaw University, DESY F35D-96-16 (1996).
%
\bibitem{HERWIG}
B.R. Webber, Ann.~Rev.~Nucl.~Part.~Sci. { 36} (1986) 253; \\
G.~Marchesini et al., Comput.~Phys.~Comm.~{ 67}~(1992)~465.
%
\bibitem{UA4}
UA4 Collab., D.Bernard et al., Phys.~Lett.~B166 (1986) 459.
%
\bibitem{ZEUS-pt}
ZEUS Collab., M.Derrick et al.,  Z.~Phys.~C67 (1995) 227.
%
\bibitem{had-mult-pt}
UA5 Collab., R.E.Ansorge et al., Z.~Phys.~C 43 (1989) 357;\\
UA1 Collab., C.Albajar et al., Nucl.~Phys.~{ B335} (1990) 261.
%
\bibitem{GRV}
M.~Gl\"uck, E.~Reya and A.~Vogt, Phys.~Rev.~{ D45}~(1992) 3986.
%
\bibitem{MRS}
A.~D.~Martin, W.~J.~Stirling and R.~G.~Roberts, 
Phys.~Lett.~{ B306} (1993) 145.
%
\bibitem{PYTHIA} 
T.~Sj\"ostrand, Z.~Phys. { C42} (1989) 301;         \\
H-U.~Bengtsson and T.~Sj\"ostrand,~Comput.~Phys.~Commun.~{ 46}~(1987)~43;\\ 
T.~Sj\"ostrand, CERN-TH.~{ 6488/92} (1992).
%
\bibitem{ZEUS-DIS-diff}
ZEUS Collab., M.Derrick et al.,  Z.~Phys.~C70 (1996) 391.
%
\bibitem{CDF-sigtot}
CDF Collab., F.Abe et al., Phys.~Rev.~D 50 (1994) 5550.
%
\end{thebibliography}
\end{document}